\documentstyle[aps,twocolumn]{revtex}
\begin{document}
\author{Osame Kinouchi \thanks{%
Electronic address: osame@dfm.ffclrp.usp.br}}
\address{Departamento de F\'{\i}sica e Matem\'atica,\\
Faculdade de Filosofia, Ci\^encias e Letras de Ribeir\~ao Preto\\
Universidade de S\~ao Paulo \\
Av. Bandeirantes, 3900, CEP 14040-901, Ribeir\~ao Preto, SP, Brazil\\
Carmen P. C. Prado\thanks{%
Electronic address: prado@fge.if.usp.br}\\
Departamento de F\'{\i}sica Geral, Instituto de F\'{\i}sica\\
Universidade de S\~ao Paulo\\
Caixa Postal 66318, CEP 05315-970, S\~ao Paulo, SP, Brazil}
\title{On the robustness of scale invariance in SOC models}
\maketitle

\begin{abstract}
A random-neighbor extremal stick-slip model is introduced. In the
thermodynamic limit, the distribution of states has a simple analytical form
and the mean avalanche size, as a function of the coupling parameter, is
exactly calculable. The system is critical only at a special point $J_c$ in
coupling parameter space. However, the critical region around this point,
where approximate scale invariance holds, is very large, suggesting a
mechanism for explaining the ubiquity of power laws in Nature. \bigskip

PACS number(s): 05.40.+j, 05.70.Ln, 64.60.Lx, 91.30.Bi.
\end{abstract}

%@twocolumnfalse\endcsname

\bigskip %]

\section{Introduction}

Self-organized criticality (SOC) is an intriguing concept which started a
large `avalanche' of research on mechanisms leading to scale invariance in
extended dynamical systems \cite{Bak}. However, there is no general
agreement about ingredients necessary to create the self-organized critical
state. This fact is reflected in the doubts about whether locally
dissipative systems really present SOC or have only a very strong divergence
of the mean avalanche size $\bar{s}$ when approaching the conservative
limit. The recent results by Chabanol and Hakin \cite{CH}, Br\"{o}ck and
Grassberger \cite{BG} and Kinouchi {\em et al.} \cite{KPP} stating that the
random-neighbor OFC model is not critical in the dissipative regime and
contradicting previous claims \cite{LJ}, is a clear example of the
difficulty of making such distinction solely on the basis of simulations. It
is also worth remembering that the prototypical sandpile (BTW) model is not
critical in the presence of local dissipation \cite{Bak,VZ,DVZ}.

The distinction between conservative/dissipative local dynamics, however, is
not what is relevant for predicting critical behavior. The decisive
ingredient seems to be the value of the coupling parameter $J$ (or the
nature of the distribution $p(J)$ in non-homogeneous systems). For example,
the Feder and Feder model with $k$ neighbors is non-conservative but is
critical when the coupling constant is equal to $J_c=1/k$ \cite{BG,FF}.

In this paper, a model is proposed which is similar to, but simpler than,
the random-neighbor stick-slip models studied in \cite{CH,BG}. For this
model, the stationary distribution of states $p_\infty (E)$ and the mean
avalanche size $\bar{s}$, as functions of the coupling parameter $J$, have
simple analytical forms (in the limit of infinite system size). The analysis
in terms of branching processes is transparent and gives a clear mechanism
for the emergence of very large but finite $\bar{s}$ in a non-negligible
region of the parameter space. In another words, although true criticality
occurs only at a special point $J_c$, there exist a large region where power
laws over several decades appear. In this region the behavior of the system
can be considered almost critical.

This occurs because the original parameter which controls the critical
behavior (the branching rate $\sigma $ in a branching process) is now, in
SOC models, a slow dynamical variable $\sigma _{t}(J)$ that depends on the
coupling parameter $J$. In our model, the stationary value $\sigma _{\infty
}(J)$ shows a plateau near the critical value $\sigma _{c}=1$, thus
enlarging the region in $J$ space where the system displays a critical
behavior. We will say that the system is critical for $J=J_{c\text{ }}$ when 
$\sigma =1$ and is 'quasi-critical 'or 'almost critical ' for values of $J$ 
where $\sigma \sim 1$. This fact may be relevant as an explanation for the
ubiquity of approximate scale invariance in nature\cite{ABLM}.

The remainder of the paper is organized as follows: In Sec. II, the model is
introduced and the main results obtained. The issue of robustness in SOC
models is discussed in Sec. III. Sec. IV contains concluding remarks and
suggestions for future work.

\section{Extremal Feder and Feder model \newline
(EFF model)}

\subsection{The model}

The EFF model is a random-neighbor version of the Feder and Feder model \cite
{BG,FF} using an extremal dynamics similar to the Bak-Sneppen model \cite{BS}%
. The extremal dynamics, that in this case substitutes (and plays the same
role that) the slow driving of the original Feder and Feder model, is here
an essential ingredient for the observation of self-organized criticality.

All sites $j=1,\ldots ,N$ have a continuous state variable $E_j\in {\cal R}$%
. At each time step the site with maximal value `fires', resetting its value
to zero plus a noise term $\eta $. Then, $k$ random `neighbors' ($rn$) of
the firing site have their states incremented by a constant $J$ plus a noise
term. The choice of neighbors is done at the firing instant: the randomness
is {\em annealed\/}. So, denoting the extremal value at instant $t$ as $%
E_i^{*}\equiv \mbox{max}\{E_j\}$, the update rules are: 
\begin{eqnarray}
E_i^{*}(t+1) &=& \eta(t) , \\
E_{rn}(t+1) &=& E_{rn}(t)+ J +\eta _{rn}(t),  \nonumber
\end{eqnarray}
with $\eta$ and $\eta _{rn}$ being random variables uniformly distributed in
the interval $[0,\epsilon]$ (the range of $\epsilon$ will be discussed
later). Note that each random neighbor receives a different quantity $%
\eta_{rn}$.

Consider the instantaneous density of states $p_t(E)$. It is clear that for
any $E$ outside the intervals $I_n\equiv [(n-1)J,(n-1)J+n\epsilon
],n=1,2,\ldots $, this density decays to zero for long times. These
intervals effectively discretize the phase space, so it is useful to define
the following quantities, 
\begin{equation}
P_n=\int_{(n-1)J}^{(n-1)J+n\epsilon }p(E)\, dE,
\end{equation}
with $n=1,2,\ldots ,n_{max}$, and $\epsilon <J/n_{max}$ so that the
intervals do not overlap (the integer $n_{max}$ will be obtained later). The
process can be thought of as a transference of sites between the intervals $%
I_n$. At each time step, one site is transferred to the interval $I_1$ and,
with probability $kP_1$, one site is removed from this interval. The average
flux to the intervals $I_n$ with $n>1$ corresponds to the probability $%
kP_{n-1}$ that a neighbor is chosen in the previous $I_{n-1}$ interval minus
the probability $kP_n$ that a neighbor is chosen in the interval $I_n$. The
average number of sites in each interval is $N_n(t)=NP_n(t)$. For long
times, that is, when the density of states outside the $I_n$ intervals goes
to zero, one can write 
\begin{eqnarray}
P_1(t+1) &=&P_1(t)+\frac 1N\left[ 1-kP_1(t)\right] , \\
P_n(t+1) &=&P_n(t)+\frac 1N\left[ kP_{n-1}(t)-kP_n(t)\right] .  \nonumber
\end{eqnarray}
Here, each time step is equal to the update of the maximal site and $k$
random neighbors.

The condition for steady states, $P_n(t+1)=P_n(t)=P_n^{*}$, gives 
\begin{eqnarray}
P_1^{*} &=&1/k,  \nonumber \\
P_n^{*} &=&P_{n-1}^{*},
\end{eqnarray}
that is, $P_n^{*}=1/k$ for all $n$. But since $p(E)$ is normalized, only $%
n_{max}$ intervals with $P_n$ of ${\cal O}(1)$ can exist. That is, 
\begin{equation}
\sum_{n=1}^{n_{max}}P_n^{*}=n_{max}\times \frac 1k=1.
\end{equation}
giving that $n_{\max }=k$.This means that $p_\infty (E)$ is composed of $k$
bumps ($n=1,\ldots ,n_{max}=k$) and the previous condition for producing
non-overlapping intervals $I_n$ reads $\epsilon <J/k$. There is also a bump
of ${\cal O}(\log N/N)$ (by analogy with the results from \cite{Flyv})
situated at the interval $I_{k+1}=[kJ,kJ+(k+1)\epsilon ]$. The other
intervals $n>k+1$ have $P_n$ of yet smaller order (see Fig.~1).

\subsection{Avalanches}

An avalanche will be defined as the number of firing sites until an extremal
site value falls bellow the threshold $E_{th} = 1$ \cite{foot}. Note that
the first site of an avalanche (the `seed') always has $E<1$ but it counts
as a firing site. So, if a seed produces no supra-threshold sites
(`descendants'), this counts as an avalanche of size one. This definition of
avalanches agrees with that used in the studies of relaxation oscillator
models.

In these random neighbor models, an avalanche can be identified as a
branching process where an active site produces $k$ new sites, each one
having a probability $p$ of being active (a `branch') and a probability $1-p$
of being inactive (a `leaf'). The branching rate $\sigma=kp$ measures the
probability that a firing site produces another firing site.

A known result for a process with a constant branching rate $\sigma $ is the
distribution of avalanche sizes \cite{BG}, 
\begin{equation}
P(s)=\frac 1s\left( 
\begin{array}{c}
ks \\ 
s-1
\end{array}
\right) \left( 1-\frac \sigma k\right) ^{ks-(s-1)}\left( \frac \sigma k%
\right) ^{s-1},  \label{ps}
\end{equation}
which, for large $s$ and small $\delta=1-\sigma$ has the form 
\begin{eqnarray}
P(s) & \approx & \frac{1}{\sqrt{2\pi(1-1/k)}} s^{-3/2}\exp(-s/s_\xi), \\
s_\xi &\approx & \frac{2(k-1)}{k}(1-\sigma)^{-2} .  \label{exponent}
\end{eqnarray}
We will see that Eq.~(\ref{ps}) can be applied to the EFF model with the
stationary value $\sigma _\infty (J)$.

Now, consider an avalanche which has terminated after $s$ sites have fired.
This avalanche is composed of one seed and $s-1$ descendants. But the
average number of descendants produced by $s$ firing sites is $\sigma s$.
Thus, on average, the relation 
\begin{equation}
\bar{s}-1=\sigma \,\bar{s},
\end{equation}
must hold, which leads to 
\begin{equation}
\bar{s}=\frac{1}{1-\sigma}.  \label{smed}
\end{equation}

Of course, this result can be obtained directly from Eq.~(\ref{ps}) after
some work. Note that $\sigma_\infty \equiv \sigma(t\rightarrow\infty)$
refers to the stationary value of the branching rate: during the transient, $%
\sigma_t $ changes with the avalanche time $t$. Although questioned by some
authors \cite{VZ}, we retain the name {\em self-organization} for this
evolution of $\sigma_t$ toward $\sigma_\infty$ mainly as a label to
distinguish these systems from standard branching processes where $\sigma$
is fixed {\em a priori\/}.

\subsection{The $J=1/k$ case}

In the case $J=1/k$, the calculation of $\sigma _\infty $ is trivial. The $k$%
-th bump, ($n=n_{\max }$) which starts at $(k-1)J$, must lie bellow the
threshold $E_{th}=1$ (if not, the system is supercritical). Then, $\epsilon $
must satisfy the condition $(k-1)/k+k\epsilon <1$, that is, 
\begin{equation}
\epsilon <1/k^2.
\end{equation}
For the standard $k=4$ neighbor case this reads $\epsilon <0.0625$. This
condition also implies that neighbors pertaining to the other bumps do not
contribute to $\sigma _\infty $, that is, cannot fire when receiving a
maximal contribution $J+\epsilon $. Now, since all the neighbors pertaining
to the $k$-th bump receive at least the quantity $J=1/k$, they are always
transformed into active sites. Thus, the average number of descendants of a
firing site is 
\begin{equation}
\sigma _\infty =kP_k^{*}=k\times \frac{1}{k} = 1,
\end{equation}
which corresponds to a critical branching process. It is known that in this
case the system presents an infinite $\bar{s}$ (see Eq.~(\ref{smed})) and,
for large $s$, a pure power law 
\begin{equation}
P(s)=\frac 1{\sqrt{2\pi (1-1/k)}}\: s^{-3/2}
\end{equation}
for the distribution of avalanche sizes \cite{BG}.

\subsection{Results for general $J$}

For the case $J<1/k$, in order to obtain an expression for $\sigma_\infty(J)$%
, the knowledge of the distribution of states $p_\infty(E)$ is required. But
it is clear that if $kJ = 1- \delta$ then inevitably $\sigma_\infty<1$ (even
for very small $\delta>0$), since some sites pertaining to the $k$-th bump
may not receive a sufficient contribution to make them active (see Eq.~(\ref
{sig2prime}) below). Thus, any value $J<J_c=1/k$ is sub-critical. This is a
common feature of many models with SOC \cite{BG,VZ,DEB}.

In our model, the calculation of $p_\infty (E)$ is very simple. In the
stationary state, a site pertaining to the $n$-th bump has energy $%
E=(n-1)J+z_n$, where $z_n$ is the sum of $n$ random variables uniformly
distributed in the interval $[0,\epsilon ]$. The distribution $p(z_n)$ may
be calculated from 
\begin{eqnarray}
p(z_1) &=&\epsilon ^{-1}\Theta (z_1)\Theta (\epsilon -z_1),  \nonumber \\
p(z_{n+1}) &=&\int_{-\infty }^\infty dz_n dz_1\,p(z_n)p(z_1)\delta
(z_n+z_1-z_{n+1}).  \nonumber
\end{eqnarray}
For the $k=4$ case, 
\begin{eqnarray}
p(z_2)&=&\epsilon^{-2}\,z_2\Theta (z_2)\,\Theta (\epsilon -z_2) 
\nonumber \\
&+&(2\epsilon -z_2)\,\Theta (z_2-\epsilon )\,\Theta (2\epsilon -z_2),
\label{pE} \\
p(z_3)&=&\epsilon ^{-3}\frac{z_3^2}2\Theta \,(z_3)\,\Theta (\epsilon -z_3) 
\nonumber \\
&+&\left( -z_3^2+3\epsilon z_3-\frac{3\epsilon ^2}2\right) \Theta
(z_3-\epsilon )\,\Theta (2\epsilon -z_3)  \nonumber \\
& +&\left( \frac{z_3^2}2-3\epsilon z_3+\frac{9\epsilon^2}2\right) \Theta
(z_3-2\epsilon )\,\Theta (3\epsilon -z_3),  \nonumber \\
p(z_4)&=&\epsilon ^{-4}\frac{z_4^3}6\Theta (z_4)\Theta (\epsilon -z_4) 
\nonumber \\
&+ & \left( -\frac{z_4^3}2+2\epsilon z_4^2-2\epsilon ^2z_4+\frac{2\epsilon^3%
} 3\right) \Theta (z_4-\epsilon )\Theta (2\epsilon -z_4)  \nonumber \\
&+ &\left( -\frac{x^3}3+2\epsilon x^2-2\epsilon ^2x+\frac{2\epsilon^3} 3%
\right) \Theta (z_4-2\epsilon )\Theta (3\epsilon -z_4)  \nonumber \\
&+ &\frac{x^3}6\Theta (z_4-3\epsilon )\Theta (4\epsilon -z_4),
\end{eqnarray}
with the shorthand $x\equiv (4\epsilon -z_4)$. The distribution $p_\infty
(E) $ has $k$ bumps. Each bump (labeled by $n$) starts at $E_n=(n-1)J$,
being proportional to $p(z_n)$ (the constant of proportionality is just $1/k$%
). In Fig.~1, the distribution $p_\infty (E)$ is compared with simulation
results for a system with $10^4$ sites, $J=0.235$, $\epsilon =0.05$ and a
sufficient number of avalanches.

For such large systems, we must be careful about using reliable random
neighbor generators. In order to speed up the search for the extremal site,
we used the binary rooted tree algorithm described by Grassberger \cite
{Grassberger95}. For example, if the system has $2^m$ sites, a binary tree
with $m+1$ levels is created such that, in each node at level $l$, it is
stored the largest value of $E$ of the two branch nodes of the $l+1$-th
level. So, the $0$-th (root) level contains the value of the extremal site.
Ascending the tree, we locate the position of this site in the upper level.
After the extremal site firing, the tree must be updated. The same occurs
when the random neighbors are updated. These operations have a complexity $%
{\cal {O}(\log N)}$ instead of the ${\cal {O}(N)}$ complexity of the naive
search mechanism.

The stationary branching rate $\sigma _\infty $ is calculated as follows.
All the sites that can be activated pertain to the $k$-th bump. When hit,
sites with $E>1-J$ are always activated. In terms of the re-scaled variable $%
z_k=E-(k-1)J$, this condition refers to sites with $z_k>\delta \equiv 1-kJ$.
They contribute to the branching rate with the quantity $\sigma ^{\prime }$, 
\begin{equation}
\sigma ^{\prime }\equiv k\int_{1-J}^1p(E) \, dE=\int_\delta ^{\delta
+J}p(z)\,dz,
\end{equation}
where $z\equiv z_k$.

Sites with $E<1-J-\epsilon $ cannot be activated and do not contribute to $%
\sigma $. Sites with $1-J-\epsilon <E<1-J$ can be activated if they receive
a quantity $J+\eta >1-E,$ that is, $\eta >\delta -z$. This occurs with
probability $P(\eta >\delta -z)=1-(\delta -z)/\epsilon $. Thus, these sites
contribute to the branching rate with the quantity 
\begin{eqnarray}
\sigma^{\prime \prime } &\equiv & k\int_{1-kJ-\epsilon}^{1 kJ}P(E)\,P(\eta
>1-E-J)\,\,dE,  \nonumber  \label{sig2prime} \\
&=&\int_{\delta -\epsilon }^\delta p(z)\,\left( 1-\frac{\delta -z}\epsilon
\right) \,dz.
\end{eqnarray}

The total branching rate is then 
\begin{eqnarray}
\sigma _\infty &=&\sigma ^{\prime }+\sigma ^{\prime \prime
}=1-\int_0^{\delta -\epsilon }p(z)\,dz  \nonumber \\
& -& \frac \delta \epsilon \int_{\delta -\epsilon }^\delta p(z)\;dz+\frac 1%
\epsilon \int_{\delta -\epsilon }^\delta z\;p(z)\,dz,  \label{sigtotal}
\end{eqnarray}
where we used the fact that $\int_0^{\delta +J}p(z)dz=1$. Since $p(z)$ has a
simple piece-wise polynomial form (see Eq.~(\ref{pE})) the

calculation of $\bar{s}$ is straightforward and the result is presented in
Fig.~2 along with simulation results for the $k=4$, $\epsilon =0.05$, for
systems with up to $N=2^{18}=262\,144$ sites. In Fig.~3, we plot simulation
results for the $P(s)$ distribution which agree very well with Eq.~(\ref{ps}%
) if $\sigma =\sigma _\infty (J)$ is used in that expression. Strong finite
size effects, however, are present when $J>0.235$.

For $\delta <\epsilon $, that is, $J_c-J<\epsilon /k$, the form assumed by $%
\sigma _\infty $ is particularly simple, since $p(z)=C\epsilon ^{-k}z^{k-1}$
in that interval ($C$ is a numerical constant). Then, 
\begin{eqnarray}
\sigma _\infty &=&1-\frac C{\epsilon ^{k+1}}\int_0^\delta z^{k-1}(\delta
-z)dz  \nonumber \\
&=&1-\frac C{k(k+1)}\left( \frac \delta \epsilon \right)^{k+1}.
\end{eqnarray}
the avalanche cutoff lenght

Since $\delta \equiv 1-kJ=k(J_c-J)$, we obtain, from Eqs.(~\ref{exponent})
and~(\ref{smed}), the avalanche cutoff size and the average avalanche size 
\begin{eqnarray}
s_\xi & = & \frac{2(k-1)(k+1)^2 \epsilon^{2(k+1)}}{C^2 k^{2k+1}}
(J_c-J)^{-\nu} , \\
\bar{s}&=&\frac{(k+1)\epsilon^{k+1}}{Ck^k}(J_c-J)^{-\nu/2}.  \nonumber
\label{sm}
\end{eqnarray}
with the critical exponent 
\begin{equation}
\nu = 2(k+1) \:.
\end{equation}
For example, with $k=4$ (which means $C=1/6$, see Eq.~(\ref{pE})) and $%
\epsilon =0.05$, the mean avalanche size is $\bar{s}=120$ already for $%
J=0.2375$. Curiously, this behavior is similar to the $\bar{s}\propto
(J_c-J)^{-k}$ divergence found in the standard random neighbor FF model \cite
{BG}.

\subsection{The EFF model with noiseless couplings}

It is instructive to compare the above behavior with that of a simpler EFF
model \cite{KPP} where the firing rule is the same, $E_{i}^{*}(t+1)=\eta \in
[0,\epsilon ]$, but the coupling between sites is noiseless, $%
E_{rn}(t+1)=E_{rn}(t)+J$. Thus, $p_{\infty }(E)$ assumes the form of $k$
rectangular bumps with $p(z_{n})=\epsilon ^{-1}\Theta (z_{n})\Theta
(\epsilon -z_{n})$. In this {\em noiseless\/} EFF model, the branching rate,
the cutoff size and the average avalanche size are 
\begin{eqnarray}
\sigma _{\infty } &=&\left\{ 
\begin{array}{cl}
0 & \mbox{for $\delta> \epsilon$} \\ 
1-\delta /\epsilon  & \mbox{for $0<\delta<\epsilon$}
\end{array}
\right. ,  \nonumber \\
s_{\xi } &=&2(k-1)\frac{\epsilon ^{2}}{k^{3}}(J_{c}-J)^{-2}  \nonumber \\
\bar{s} &=&\frac{\epsilon }{k}\left( J_{c}-J\right) ^{-1}.
\end{eqnarray}
In contrast with the noisy model, large avalanches only occur when $J$ is
very close to $J_{c}$ (see Fig.~2). Thus, the EFF model with noiseless
couplings does not present an enlargement of the region where the system
displays a critical behavior as observed in the noisy EFF model.

\section{On SOC definitions}

The idea of self-organized criticality present in the literature embodies
two distinct properties. The term {\em critical\/} refers to the existence
of power laws and to the absence of a characteristic scale in the response
of the system to the driving mechanism of the dynamics; the term {\em %
self-organized\/} refers to the fact that there exist a parameter ($\sigma_t$%
), which controls the avalanching process, whose value is not fixed {\em a
priori\/} like, for example, in standard percolation and branching
processes. This parameter evolves in time, during a transient phase, toward
a stationary value $\sigma_\infty $. Indeed, this time dependence should be
written as $\sigma_t =\sigma(p_t(E))$, that is, $\sigma_t $ is a functional
of the distribution of states $p_t(E)$, that, in turn, evolves toward a
statistically stationary distribution $p_\infty(E)$. So, $\sigma_\infty
\equiv \sigma(p_\infty(E))$. If $\sigma_\infty = \sigma_c=1$, the system is
critical.

The evolution of $p_t(E)$ toward the steady-state $p_\infty (E)$ is akin to
the transient relaxation in equilibrium systems: any initial condition leads
to the same stationary state, thus to the same value of $\sigma _\infty $.
However, this robustness to initial conditions and external perturbations on 
$p(E)$ (`dynamical stability') should not be mistaken as parameter
robustness (`structural stability'). This is a distinct characteristic
claimed to be present on some SOC models (see for instance \cite
{Bak,Grinstein,OFC,BD}). For a system to have `structural stable
criticality', there would be a finite parameter range for which, after the
transient, the system is critical. In this case, $\sigma _\infty (J)=\sigma
_c$ for $J$ belonging to some interval $[J_c,1/k]$. That kind of behavior
will be also called by us {\em generic SOC\/}.

`Structural stability' is a relative concept which depends on the parameter
space physically available for the system. For example, it is well known
that the sandpile model is not critical in the presence of dissipation. The
sandpile dissipation parameter corresponds to the quantity $\delta =1-kJ$ in
our model \cite{VZ,DVZ}. The standard BTW model is by definition `tuned'
into a critical state through the `imposition' of a conservation law.
Although it could be argued that dissipation is not a natural feature of
sandpiles, since sand does not disappear, the appearence of SOC in nature
would sound much more natural if criticality could be observed over a region
of the parameter space, not only in a special point.

Generic self-organized criticality is depicted in curve~(a) of figure 4. In
this case, there is a finite range of $J$ values for which $\sigma _{\infty }
$ assumes the critical value $\sigma _{c}=1$. In this figure, curves (d) and
(e) represent the behavior observed in the BTW model and also in the
noiseless EFF model examined above, for which the system is critical only
for a special value of the parameter $J$. However, there is a third
possibility. Curves (b) and (c) represent the behavior of $\sigma (J)$ given
by Eq.~(\ref{sigtotal}) for the EFF model with noisy couplings: although the
system is critical only at $J=J_{c}$, the system is `almost critical' over a
large parameter region. This behavior has also been observed in the standard
random neighbor versions of FF and OFC models \cite{BG}. The importance of
this characterization is that several models in the SOC literature,
previously seen as having true generic criticality, are now recognized as
having only an almost critical behavior as discussed above.

A model which apparently presents generic SOC behavior in coupling space is
the two-dimensional OFC model \cite{OFC,Grass,PCDCA}. Also the standard
Feder and Feder model \cite{FF} is claimed to be critical for $J<J_c$ \cite
{BD,PCDCA}. Looking at the behavior of the models studied so far, we make
the following conjecture: a necessary condition for a lattice model to
present a generic SOC behavior is that its corresponding random neighbor
version already presents an enlarged critical region in the sense discussed
above. This could be tested by comparing the 2D versions of the EFF and
noiseless EFF models studied above.

In conclusion, we found that some systems that display SOC, although being
critical only for a single value for $J$, are almost critical region in a
large region of the parameter space. This almost critical behavior is
difficult to be distinguished, in practice, from true generic SOC behavior:
both in numerical simulations (huge lattices would have to be used) and in
Nature (due to limitations in the data) power laws can only be measured over
some scale decades \cite{ABLM}. So, in order to explain the ubiquity of
scale invariance in Nature, having a true generic SOC or only presenting an
enlarged region where the system is almost critical are, as far one can
measure, identical.

\section{Conclusions}

A class of extremal stick-slip models has been introduced and studied in the 
$N\rightarrow \infty $ limit. We showed that noise in the couplings of the
EFF model changes the exponent that controls the amplitude of the critical
region from $\nu=2$ to $\nu=2(k+1)$. This enlargement of the region where
the system displays a critical behavior is similar to that found in the
standard random neighbor OFC and FF models \cite{CH,BG,KPP}. Like in other
models, the true critical state occurs only for one point in parameter space 
\cite{CH,BG,VZ,DVZ,DEB}, but in practice that fact can hardly be noticed,
and the model displays the typical features of generic SOC.

In future work we hope to determine the minimal ingredients for producing
the enlargement of the critical region in the models examined in the SOC
literature. We will also present results for the two-dimensional case and
compare with the standard OFC and FF models. The simple mechanism devised in
this work suggests that, if true generic criticality is not easy to obtain
in the space of possible models, this quasi-critical behavior certainly is.
Thus, for explaining the robustness of approximate scale invariance in
Nature, this mechanism seems to be more ''generic'' than generic criticality.

{\bf Acknowledgments:} The authors thank P. Bak, S. R. A. Salinas, Suani T.
R. Pinho for helpful discussions, N. Dhar for remarks about the SOC concept
and K. Christensen, R. Dickman, J. F. Fontanari, Nestor Caticha, D. Alves
and R. Vicente for commenting the manuscript. O.K. thanks FAPESP for
financial support.

\pagebreak

{\Large {\bf FIGURE CAPTIONS}}

\bigskip

{\bf Figure 1:} Distribution of states $p_\infty (E)$ for $k=4$, $J=0.235$
and $\epsilon =0.05$: theoretical (solid) and simulation (circles) with $N =
10^4$ sites.

{\bf Figure 2:} Mean avalanche size $\bar{s}$ as a function of parameter $J$%
. Theoretical (solid) and simulations with $N$ up to $2^{18}=262\,144$ sites
for noisy EFF model (circles) with $\epsilon =0.05$ and noiseless EFF
model(triangles) with $\epsilon =0.2$. These $\epsilon $ values are chosen
such that the last interval ($I_4$) has the same length in both models.

{\bf Figure 3:} Simulation results ($N=2^{13}=8192$ sites, $k=4,\epsilon
=0.05$) for the distribution $P(s)$ with $J=0.21,0.22,0.23,0.235$ (from left
to right), compared with theoretical curves (solid).

{\bf Figure 4:} a) Generic self-organized criticality: the value of
parameter $\sigma _\infty $ is critical on a finite range of the system
parameter $J$; b)$\;\epsilon=0.0625$ and c)$\;\epsilon=0.05$, enlargement of
the critical region (EFF model with noisy couplings, $k=4$): $\sigma_\infty$
is almost constant near $J_c $; d)$\;\epsilon=0.25 = 4 \times 0.0625$ and e)$%
\;\epsilon=0.2 = 4 \times 0.05$, standard critical behavior (EFF model with
noiseless couplings): the coupling parameter $J$ must be very close to 0.25
for obtaining $\sigma _\infty \approx \sigma _c$ due to the linear behavior
of $\sigma_\infty(J)$. Note that, in the noiseless couplings case, $\epsilon$
refers to the amplitude of the noise received by the extremal site after
discharge.


\begin{references}
\bibitem{foot}  This special choive for the threshold has no influence on
the results, since $E_{th}$ only defines the scale of the $E$ axis.
\end{references}
\end{document}